\documentclass[sigconf]{acmart}

\AtBeginDocument{%
  \providecommand\BibTeX{{%
    \normalfont B\kern-0.5em{\scshape i\kern-0.25em b}\kern-0.8em\TeX}}}

\usepackage{subcaption}
\usepackage{float}
\usepackage{natbib}
\def\citepos#1{{\hypersetup{citecolor=black}\citeauthor{#1}}'s \citep{#1}}

\let\oldciteauthor=\citeauthor
\def\citeauthor#1{{\hypersetup{citecolor=black}\oldciteauthor{#1}}}

\copyrightyear{2022}
\acmYear{2022}
\setcopyright{acmlicensed}\acmConference[CHI '22]{CHI Conference on Human Factors in Computing Systems}{April 29-May 5, 2022}{New Orleans, LA, USA}
\acmBooktitle{CHI Conference on Human Factors in Computing Systems (CHI '22), April 29-May 5, 2022, New Orleans, LA, USA}
\acmPrice{15.00}
\acmDOI{10.1145/3491102.3502124}
\acmISBN{978-1-4503-9157-3/22/04}

\begin{document}

\title[How Interest-Driven Content Creation Shapes Opportunities for Informal Learning in Scratch]{How Interest-Driven Content Creation Shapes Opportunities for Informal Learning in Scratch: A Case Study on Novices’ Use of Data Structures}

\author{Ruijia Cheng}
\orcid{0000-0002-2377-9550}
\affiliation{%
  \institution{University of Washington}
  \city{Seattle}
  \state{Washington}
  \country{USA}
  \postcode{98195}}
\email{rcheng6@uw.edu}

\author{Sayamindu Dasgupta}
\orcid{0000-0001-6083-2114}
\authornote{Dasgupta was at the University of North Carolina at Chapel Hill when this work was submitted for review. Since then, he has moved to the University of Washington.}
\affiliation{%
  \institution{University of North Carolina at Chapel Hill}
  \city{Chapel Hill}
  \state{North Carolina}
  \country{USA}
 \postcode{27599}}
\email{sayamindu@unc.edu}

\author{Benjamin Mako Hill}
\orcid{0000-0001-8588-7429}
\affiliation{%
  \institution{University of Washington}
  \city{Seattle}
  \state{Washington}
  \country{USA}
  \postcode{98195}}
\email{makohill@uw.edu}

\renewcommand{\shortauthors}{Cheng, R., et al.}

\begin{abstract}

Through a mixed-method analysis of data from Scratch, we examine how novices learn to program with simple data structures by using community-produced learning resources. First, we present a qualitative study that describes how community-produced learning resources create archetypes that shape exploration and may disadvantage some with less common interests. In a second quantitative study, we find broad support for this dynamic in several hypothesis tests. Our findings identify a social feedback loop that we argue could limit sources of inspiration, pose barriers to broadening participation, and confine learners' understanding of general concepts. We conclude by suggesting several approaches that may mitigate these dynamics.
\end{abstract}

\ccsdesc[500]{Human-centered computing~Empirical studies in HCI}
\ccsdesc[500]{Human-centered computing~Empirical studies in collaborative and social computing}
\ccsdesc[500]{Human-centered computing~Collaborative and social computing theory, concepts and paradi\-gms}
\ccsdesc[300]{Human-centered computing~Computer supported cooperative work}
\ccsdesc[500]{Social and professional topics~Computing education}

\keywords{online communities, informal learning, learning to code, connected learning, computational participation, computers and children, creativity support tools, social computing and social navigation, forums, Scratch}

\maketitle

\section{Introduction}

Scholars increasingly look to interest-driven online communities as promising environments for supporting learning \citep{bruckman_community_1998, ito_hanging_2009, jenkins_confronting_2009}. These communities rely on user engagement in content creation to curate community-produced learning resources where users engage in sharing artifacts that they create and online discussion with other users. 
Although such communities exist in a range of domains like creative writing \cite{campbell_thousands_2016}, graphical design \cite{marlow_rookie_2014}, and more, many of the largest and most studied have focused on supporting young people in computational learning (i.e., learning about computational concepts, often through learning to program a computer). Widely studied examples include the Scratch community \cite{monroy-hernandez_scratchr:_2007} and MOOSE Crossing \cite{resnick1998moose}. These interest-driven computational learning communities are built around the idea of ``computational participation'' \cite{kafai_connected_2014}, which encourages learners to develop programming skills through creating and sharing projects and interacting with other learners \cite{kafai_computational_2016, bruckman_community_1998, brennan2011more}.  

Despite this promise, it remains unclear how the creation and usage of community-generated learning resources support the learning of computational concepts. Previous studies of the Scratch online community show that users who remix, or build projects from code shared by other users, achieve better learning outcomes \cite{dasgupta_remixing_2016}. However, most users do not demonstrate much innovation in remixed projects, making people question whether they actually develop transformative abilities \cite{hill_remixing_2013}. In fact, most Scratch users only display a limited range of programming skills \citep{matias_skill_2016, yang_uncovering_2015}. For example, only around 15\% users have ever used data structures---an important computational concept---in Scratch projects \cite{dasgupta_remixing_2016}. While online discussion provides opportunities for learners to mentor each other and collaboratively debug programs \citep{fields_i_2015, shorey_hanging_2020}, it can also end up as superficial socialization in ways that can even act as a barrier to the exchanges of ideas, feedback, and resources \citep{shorey_hanging_2020}. These mixed signals suggest the need of a better understanding about the mechanism of interest-driven content creation in computational learning communities. How does computational learning happen through interest-driven content creation? What is the role of community in the process? How do community-produced learning resources support learners' diverse interests?

To explore these questions, we present two studies about the Scratch community that describe the opportunities and challenges that interest-driven content creation and related community activities introduce to computational learning. In Study 1, where we present a grounded theory analysis of 400 discussion threads in the Scratch forums about how learners develop an understanding of data structures---variables and lists. %
Based on this analysis, we hypothesize a social feedback loop where engagement in content sharing and Q\&A naturally raises the visibility of some particular ways of using variables and lists. Through their increased visibility, these examples become archetypes that can limit the breadth of future projects in the community. In Study 2, we conduct a quantitative analysis of the code corpus of more than 200,000 Scratch projects to test our hypothesis and find statistical support for the social process theorized in our first study.
We conclude with several implications for design and content curation that we believe could improve support for diverse interests in Scratch and similar interest-driven learning communities. 

This work makes several contributions to the HCI and social computing literature on computational learning. First, we make an empirical contribution by presenting detailed qualitative and quantitative evidence about how novices learn to use data structures in the Scratch online community.
Second, we offer a theoretical contribution in the form of a framework describing a dynamic process where interest-driven content creation can both assist learning about particular topics while posing important limits on the ways that those topics are engaged with.
Finally, we make a contribution to the literature on the design of informal learning systems by speculating about how online interest-driven communities can be designed to mitigate the negative repercussions of the dynamic we describe.

\section{Background}

\label{sec:background-learning}

Online communities are increasingly common settings for participatory, interest-driven, and community-supported learning \cite{bruckman_community_1998, gee_situated_2006, jenkins_confronting_2009, jenkins_participatory_2016}. 
A broad range of theoretical frameworks have been used to design and analyze these communities---many building on foundational theories on the social origins of learning by \citet{vygotsky_mind_1978} and \citet{lave_situated_1991}. Another key theoretical framing is \citet{papert_situating_1991-1}'s view of learning as the construction of knowledge that ``happens especially felicitously in a context where the learner is consciously engaged in constructing a public entity'' (p.~1) and which emphasizes the importance of interest-driven exploration and ``personally powerful ideas'' in promoting learning \cite{papert_mindstorms_1993}. Recent scholarship on connected learning \cite{ito_connected_2013-1, ito2018affinity} have also endorsed the role of shared interest and participatory culture in building learning communities. 

Learning experiences in interest-driven online communities happen through two primary pathways: through sharing creative artifacts like fan fiction \cite{campbell_thousands_2016}, design mock-ups \cite{cheng_critique_2020}, interactive computer programs \cite{dasgupta_children_2016}; and through social interactions around these artifacts like commenting, remixing, and critiquing. To promote the first pathway, many communities are structured so that the products of learning are made visible as public artifacts that can be used by others as illustrative examples and for inspiration \cite{dasgupta_remixing_2016, gan_gender_2018, marlow_rookie_2014} as well as scaffolds for replication, practice, and innovation \cite{dasgupta_remixing_2016, tausczik_share_2017}. To promote the second pathway, communities feature direct user-to-user support such as comments \cite{campbell_thousands_2016}, forum posts \cite{kou_supporting_2017}, and Q\&A discussions \cite{tausczik_collaborative_2014} that can help members gain an understanding of specific topics or techniques \cite{shorey_hanging_2020}. The two pathways are deeply interwoven. By making artifacts publicly visible, creators are able to receive constructive feedback that can support learning \cite{gan_gender_2018, yen_social_2016}. This often includes input from experts and professionals that is otherwise unavailable \cite{kou_what_2018, hui_distributed_2019} as well as social recognition and support \cite{campbell_thousands_2016}. Additionally, learners in online communities often center their social interaction around discussions of public artifacts and as social interactions support the further production of artifacts \cite{kim_mosaic:_2017}, collaborative problem-solving \cite{tausczik_collaborative_2014, li_is_2015, shorey_hanging_2020} and community-wise knowledge advancement \cite{gray_co-producing_2019}.

In recent years, creative programming communities have emerged as a prominent example of online interest-driven learning communities. In what \citet{kafai_computational_2016} has described as ``the social turn in K–12 computing'' (p.~27) and ``computational participation,'' scholars have turned to interest-driven and socially supported contexts to promote learning about computing where learners create programs to be shared with peers. Through the work of efforts inspired by this approach, millions of young people have engaged in programming in online communities.
Some notable examples of these communities include programmable multiplayer game environments \cite{bruckman_community_1998}, platforms for interactive media creation \cite{monroy-hernandez_scratchr:_2007, resnick_scratch:_2009, wolber_app_2011}, and amateur technical support groups \cite{fiesler_growing_2017}.

Despite the promise of these communities, %
it remains unclear how to best promote computational participation. Recent studies of online programming communities have shown unequal outcomes in terms of both participation and learning based on gender and race \citep{fields_programming_2014, gan_gender_2018, richard2016blind} and that important debugging or collaborative sense-making activities are not always helped by socialization \cite{shorey_hanging_2020}. Furthermore, while online communities allow users to gain inspiration from examples posted by others, studies on remixing activities indicate it can negatively impact originality \cite{hill_remixing_2013}. These examples indicate a lack of a general understanding of the dynamics around learning in online informal contexts. In interest-driven communities of all types, learning pathways can be blocked by the difficulty of ensuring high quality content \cite{kotturi_why_2019, marlow_rookie_2014, hui_distributed_2019, xu_what_2012, agichtein_finding_2008} and the challenge of engaging diverse groups of users \cite{cheng_building_2020, buechley_lilypad_2010}. This mismatch between the theoretical promise of online interest-driven communities and what is seen in practice indicates a lack of understanding of why user engagement supports or fails to support learning and how we can best design to facilitate positive learning outcomes.

Because computational participation involves learning many different concepts, we focus on learning experiences over one specific computational concept that has been the subject of substantial academic work: the simplest data structures comprising scalar variables and lists.
We consider data structures specifically because \citet{brennan_new_2012} identify the ability to understand how to store, retrieve, and update data as one of seven major practices that contribute to computational thinking \citep{barr_computational_2011, denning_computational_2019, wing_computational_2006}. Despite their importance, research has shown that data is the least commonly engaged computational thinking concept in Scratch \citep{dasgupta_remixing_2016}. Previous work by \citet{dasgupta_remixing_2016} estimates that less than 15\% of Scratch users will ever make projects using data structures. When used, it is often engaged with in superficial ways \citep{blikstein_pre-college_2018}. %

Why are data structures hard to learn? An explanation stems from the fact that learning about a computational concept involves learning its structural and functional uses \cite{disessa_models_1986}. The \textit{structural uses} of variables and lists (i.e., how to integrate them in a program) are straightforward. For example in Scratch, there are only two methods (\texttt{get()} and \texttt{set()}) that represent the structural usage of scalar variables. However, the \textit{functional uses} of variables and lists (i.e., what meaningful outcome that they can help create) are both broad and invisible. For example, while variables can be used for storing user input, keeping track of internal program state, counting in a loop, and so on, these functions may not be immediate obvious to novices. Previous work has argued that learners require  ``specific tutoring'' \citep[p.~207]{disessa_models_1986} and has demonstrated that scaffolds are required to cope with misconceptions about how variables can be used to achieve concrete functionality \cite{hermans_thinking_2018}. It is unclear where interest-driven communities like Scratch are effective in providing proper tutoring needed for learning the functional uses of data structures. Therefore, we ask the following research question: \textit{When does interest-driven content creation most effectively support learning of functional uses of variables and lists? When does it fall short?} 
We seek to answer these questions through two closely linked empirical studies that unpack practices, challenges, and opportunities for learning about variables and lists in the Scratch online community.

\section{Empirical Setting}
\label{sec:scratch}

\begin{figure*}[t]
     \centering
     \begin{subfigure}[b]{0.3\textwidth}
         \centering
         \includegraphics[width=\textwidth]{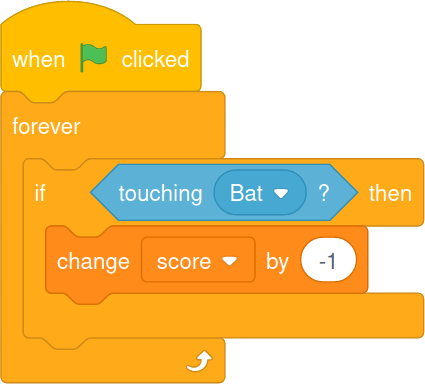}
         \caption{Sample Scratch code showing a variable in the form of a data block called ``score'' being decremented on collision with a bat sprite}
         \Description{A program made by a stack of Scratch programming blocks. From the top to bottom: ``when clicked,'' ``forever,'' ``if touching Bat? then,'' ``change score by -1.''}
         \label{fig:scratchcode}
     \end{subfigure}
     \hfill
     \begin{subfigure}[b]{0.68\textwidth}
         \centering
         \includegraphics[width=\textwidth]{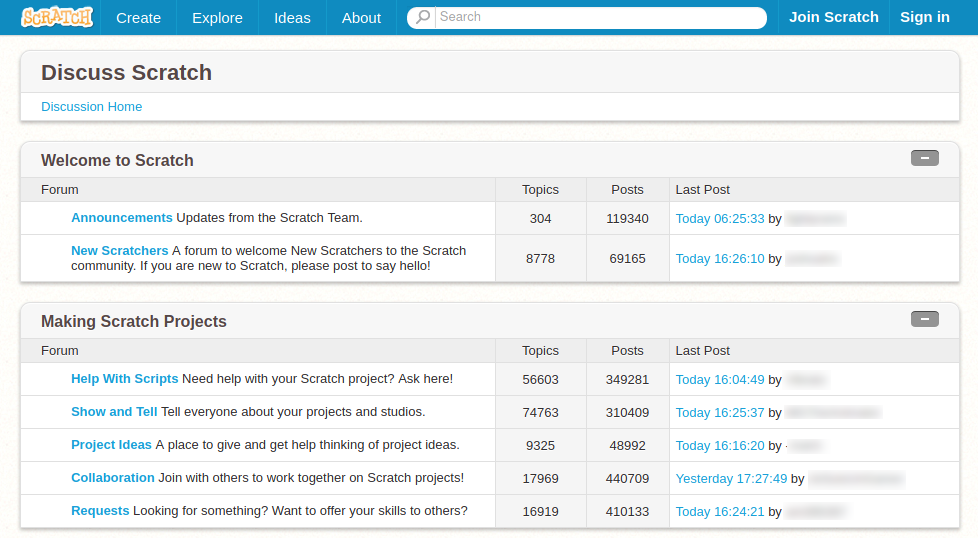}
         \caption{Index page of the Scratch forums}
         \Description{A screenshot of the index page of the Scratch forums. From top to bottom: big title ``Discuss Scratch,'' hyper link ``Discussion Home,'' Secondary title ``Welcome to Scratch,'' a list of forums in hyper links ``Announcements,'' ``New Scratchers,'' Secondary title ``Making Scratch Projects,'' a list of forums in hyper links ``Help with Scripts,''  ``Show and Tell,''  ``Project Ideas,'' ``Project Ideas,'' ``Collaboration,'' ``Requests.''}
         \label{fig:scratchforums}
     \end{subfigure}
     \caption{The Scratch programming language and online community}
\end{figure*}

Scratch is a visual, block-based programming language designed for children \cite{resnick_scratch:_2009}. Scratch programming primitives are represented by visual blocks that control the behavior of on-screen graphical objects called sprites. Scratch programs (commonly called projects) are constructed by dragging and dropping blocks together. As a programming language, Scratch supports basic data structures in the form of scalar variables and vector lists (Figure \ref{fig:scratchcode}).

Primitives to operate on variables and lists fall under the category of ``data blocks'' in Scratch and their design are described in detail by \citet{maloney_scratch_2010}. When creating a variable or list, Scratch users assign a name to the variable or list through a free-form text entry field that is invoked through a ``make a variable'' or a ``make a list'' button in the Scratch user interface. We refer to these user-defined names as ``variable name'' or ``list name'' in this paper. When referring to the the variable or list within the Scratch program, users select the appropriate name from a dropdown list embedded in the block.
Variables in Scratch have two forms which share a grammar: (1) conventional variables and lists which are local to each instance of a running project; and (2) cloud variables which are persistent across multiple executions of a project and shared across users \citep{dasgupta_surveys_2013}. Cloud variables are only accessible to established members in the Scratch online community as established by an undocumented set of criteria \citep{dasgupta_how_2018}.

Scratch is situated within an online community where anyone can sign up and share their projects, comment on, ``like,'' and bookmark others' works, and socialize in forums \cite{monroy-hernandez_scratchr:_2007}.\footnote{\url{https://scratch.mit.edu/}} As of 2021, the Scratch online community has over 65 million registered users, and over 68 million shared projects that span a diverse range of genres and themes.
The large majority of Scratch users are between 8--16 years old and the median age for new contributors is around 12.\footnote{All statistics about Scratch community activity and users are taken from the public information on: \url{https://scratch.mit.edu/statistics/}} Although our data might include adults, we follow other scholarly accounts and refer to Scratch users as ``kids'' \citep[e.g.,][]{ito_hanging_2009}.
We draw data from both the Scratch community itself and its discussion forums, shown in Figure \ref{fig:scratchforums}. These forums comprise a number of topical forums organized into categories such as ``Making Scratch Projects'' and subcategories such as ``Help With Scripts'' or ``Project Ideas'' \citep{scaffidi_how_2012}.\footnote{\url{https://scratch.mit.edu/discuss/}}

\section{Research Ethics}
\label{sec:ethics}

This research relies on two sources of data and included no intervention or interaction with subjects. Although the data in our qualitative analysis (§\ref{sec:study1}) are public posts in the Scratch forums, we sampled from these posts using keyword searches of a copy of the Scratch Forums database in a way that the public could not do easily. 
We have obscured users' identities by replacing usernames with alphanumeric identities and by following advice from \citet{markham_fabrication_2012} to reword quotes to make it more difficult to identify the Scratch users we quote using simple web searches.
Our quantitative analysis of the Scratch code corpus in §\ref{sec:study2} relied on data that the Scratch team has published as part of the \textit{Scratch Research Dataset} \citep{hill_longitudinal_2017}.
This work was reviewed and overseen by the IRB at MIT as part of a broader protocol covering observational studies of Scratch. Our institutional IRBs delegated oversight of the work on this project to the MIT IRB.

\section{Study 1: Theory development}
\label{sec:study1}

Because we could not identify existing theories with clear prediction on when Scratch would best support learning about variables and lists, we began with an open-ended interpretive analysis that sought to build a theory about kids' practices, learning, and engagement in discussions about simple data structures.

\subsection{Methods}

To build a dataset for our qualitative analysis, we generated a sample of 400 discussion threads about variables and lists from the Scratch discussion forums.
Because we were interested in how kids learn to make projects with variables and lists by engaging in learning resources curated in discussions, we limited our sample to two subforums that emphasized question asking: \textit{Help With Scripts} and \textit{Questions about Scratch}.
We chose to study forum threads instead of the more widely used project comments because previous work suggested that only a small number of comments were related to problem solving \cite{shorey_hanging_2020}.
The dataset that we used for sampling
contains discussions that took place between October 11, 2012 and April 5, 2017.

To acquaint ourselves with our setting, we spent several weeks browsing the forums. As part of this process, we found that the many posters of threads about data structure did not use the specific terminology of ``data block,'' ``variable,'' or ``list'' in their posts. Therefore, to include a broad range of conversations about data in the Scratch forums, we followed advice from \citet{trost_statistically_1986-1} to build a ``statistically non-representative stratified sample'' that would ensure that a range of different ways of talking about data were reflected in our data, but without concern for each type's prevalence.
To do so, we sampled threads in three stages using different keywords. First, we randomly sampled 100 threads from titles containing the keywords ``variable,'' ``list'' or ``data.''
Second, we sampled 10 random threads each from the 11 most common variable or list names---as identified in an analysis of the Scratch code corpus by \citet{dasgupta_children_2016}---for 110 threads total. These terms included ``high scores,'' ``lives,''  ``inventory,'' ``leaderboard,'' ``speed,'' ``timer,'' ``counter,'' ``words,'' ``points,'' ``velocity,'' and ``answers.''
Third, to further increase diversity in our dataset, we randomly sampled two threads with each of the other top 100 variable and list names. This step resulted in an additional 200 threads. We only included threads with more than a single post because we wanted to ensure that the thread contained at least some form of a discussion.
All together, these sampling steps resulted in 410 threads. Because 10 threads were included in more than one of our samples, we ended up with a total of 400 threads that contained 2,790 posts and 963,593 words of content---equivalent to 547 pages of single-spaced text.

We analyzed these data using \citepos{charmaz_constructing_2006} approach to constructing grounded theory. In the open coding phase, the first author led the analysis and started by annotating the threads with both line-by-line and incident-by-incident codes. In the process, the first author regularly shared coded data with the rest of the team and iteratively update the codes based on team discussion. For example, in vivo codes were generated about what kids were creating using data structures: ``making a score counter,'' ``making a leaderboard,'' and so on. Following Charmaz, a small number of ``sensitizing codes'' drawn from existing theories were also used in the open coding phase, such as ``engaging in collaborative debugging,'' \cite{shorey_hanging_2020} ``helping through remixing,'' \cite{dasgupta_remixing_2016} and so on. In the axial coding phase, the first author led the process of code development by grouping initial codes into themes and meta-themes. For example, the initial codes about what kids were creating using data structures were grouped into an axial code called ``functional use of data,'' with sub-level axial codes on variables and lists. This process involved repeated team discussions on code development, several rounds of iteration on code schemes, and re-coding data. Lastly, we composed memos to describe connections among the axial codes. Our findings reflect the content of the memos written about the major themes that emerged at the end of our analysis.

\subsection{Findings}
\label{sec:qualfindings}

Our analysis resulted in three major themes. First, we discovered evidence that learners, driven by an interest in making specific popular game elements, tend to adopt a narrow set of functional uses of variables and lists. Second---and as a result of the first finding---we found that user-generated learning resources about variables and lists are framed around those specific examples of functional uses. Finally, we identified that those specific examples become archetypes that restrict the breadth of future functional uses in the community. Together, these findings describe a grounded theory about how interest-driven content creation can limit learning opportunities.

\subsubsection{Learners create projects with functional uses of variables and lists specific to their interests in game-making}
\label{sec:games}

We found that Scratch users were often introduced to variables and lists when they engaged in discussions about specific functional elements of their projects. Because kids in our sample usually had specific goals for their projects in mind, but little knowledge about how to realize those goals in code, they would describe the particular thing they wanted to do when seeking help. 
The concepts of variables or lists would typically be brought up by someone else in response.
Although a quarter of our sample were not selected on variable names and half of our sample included threads based on 100 different variable names, game-making was an almost ubiquitous topic of discussion. Over and over, we observed kids phrasing their questions in terms of game-related goals in discussions in which variables and lists were eventually brought up. In the following sections, we discuss the canonical game-related use cases for variables, lists, and cloud variables in turn.

\textbf{Variables}: Two examples of common game elements that kids like to make in Scratch are \textit{score counters} and \textit{animations}. Score counters are a game element that keeps track of a player's score or remaining lives. Although this game element is broadly familiar to Scratch users, a user seeking to implement a score counter for the first time may not know about variables. Indeed, it is not always even obvious to users who know about the existence of data blocks in Scratch that a variable is an appropriate way to keep track of a changing quantity. Furthermore, it can be challenging for kids to implement a counter and integrate it into their program. For example, K1 asked: ``I would like to know how to add lives in my game. I want it so that whenever the main character touches a ghost, it would lose one life.'' %
In interactions like these, other users would introduce variables and their functional use as a score counter. In this case, a reply from K2 suggested K1 ``create a variable with the name lives''
and use it to control the visual elements that represent the character's lives. Although carefully scoped to the specific problem faced by K1, the reply highlighted the role of variables in tracking changes. 

Another common pathway to learning about variables involved animated objects in games. Animation frequently relies on variables because it involves changing the speed or size of objects when triggered by conditions. For instance, K3 asked:

\begin{quote}
I'm making a pong game where I want to add a control to tell the ball to go faster. Is there a button for this? If not, how can I make this work?
\end{quote}

\noindent K3 arrived to the Scratch forums knowing that they wanted to vary the speed of a pong ball. For them, the challenge was the specific case of making a ball move at a range of different speeds. Responding to their question, K4 pointed to variables:

\begin{quote}
Somehow, you must tell the ball to move. Make sure you use a variable. Like use ‘move [speed] steps’ rather than ‘move 5 steps’. Then you just need to set your variable to the speed you like.
\end{quote}

\noindent This response suggested that the K3 use variables and told them how. In these examples and many others, we saw that variables---both the concept and the term---almost never appeared in learners' initial questions. Instead, many of the Scratch forum's learning resources about variables existed in answers to questions about score counters, animation, and other game elements.

\textbf{Lists}: We found a similar pattern for the list data structure. Frequently, kids were introduced to list data structures when trying to figure out how to make inventories---a game element with which players can store and retrieve items. 
For example, K5 asked: ``Does anyone have an idea how to make a good inventory for a game?'' K6 answered: ``You can use the list block to store your items in the inventory.'' In another example, K7 said, ``Allllllright so I'm making an inventory for game (who wouldn't want one?) so I don't know how to make one. Can anybody help?'' These kids were all pointed to lists. Discussions like these helped kids who were struggling to build inventory features connect the abstract concept of a list with its functional use of storing multiple game items and  played out repeatedly in the forums. %

As with variables, kids imagined how an inventory would be used in the context of their games. For example, they described backpacks of weapons or a pool of correct answers in a quiz game. These kids also imagined rules describing how a player should manipulate items in the inventory. For example, K8 wanted to make a weapon inventory to hold ``basic armour'' and hoped to ``make the player lose less blood when he/she has those items.''
After sharing these ideas, they received suggestions to put a list data structure within an ``if else'' statement. Cases like this suggest that building inventories allowed kids to learn not only about how to populate and read from lists, but also about basic list operations like deleting and appending items and about conditions and loops. In some cases, more advanced learners would describe methods for accomplishing complicated tasks imagined by novices: 

\begin{quote}
K9: I am making a game where you can buy food and eat it. I want it so you can delete a certain food item from a list... so when you click, the sprite called ‘Strawberry Popsicle’ disappears, but also the name ‘Strawberry Popsicle’ disappears from the list too.

K10: You need to search in the list and find the item that you want to delete. You can just look at each item in the list and compare it to the one you are looking for. Then you stop when you find it or get to the end of the list. This is called a sequential search. [Example code to solve the problem]
\end{quote}

\noindent This thread shows how relatively sophisticated algorithms were explained in terms of very specific use cases, often with example code. By exchanging ideas about inventories in games, kids introduced each other to lists, their function, and the way they could be used.

\textbf{Cloud Variables}: A final example extends this pattern to \textit{cloud variables} (see §\ref{sec:scratch}). Cloud variables' ability to store data in ways that are persistent and shared were essential for users building  ``leaderboards'' or ``high score'' systems that could record, rank, and display scores from multiple players---e.g., ``a leaderboard in which the highest scores of every player of the game could be saved'' (K11).
Discussions about leaderboards often involve pointing out the existence of cloud variables, 
the differences between local and cloud data, and ideas about how to write code to use both. These conversations often segued into advanced programming topics like the encoding and decoding of strings. As with variables and lists, these conversations typically remained focused on the specific use case of leaderboards.

In all three cases, specific use cases became linked to specific data structures---variables with counters and animations, lists with inventories, and cloud data with leaderboards. Because questions tended to focus on these types of elements, discussions about solutions did as well. Through this process of user-to-user support, kids learned how to apply data structures in an informal and unstructured manner. As we describe in the next section, both these conversations and the games that Scratch users created acted as learning resources about variables and lists that were subsequently used by other learners in the Scratch community. 

\subsubsection{User-generated learning resources about variables and lists are framed around specific examples of functional uses}
\label{sec:case_based_qanda}

One feature of informal online learning environments is that conversations and solutions act as learning resources for subsequent participants facing similar challenges.
In ways that are visible in our examples in §\ref{sec:games}, both the questions posed and the answers provided in our sample tended to focus on specific game-related functional uses.
As a result, learning resources about how to use variables and lists tended to be framed in terms of specific game elements and rarely engaged with the more general concepts about data structures. For example, the following threads show a discussion on how to change the speed of a ball using variables:

\begin{quote}
K13: Can someone help me figure out how to change the speed of the ball when it hits the paddle a certain number of times?

K14: You can make a variable called HITS, set it to 0, change it by 1 every time you hit the paddle. When it gets to the number you like, change costume and set counter to 0. 
\end{quote}

\noindent In this exchange, K14's answer details exactly what K13 should do to solve their problem---specified down to the names of variables. While this answer likely solved K13's problem, it is anchored on a very specific functional use of variables and did not explain, or suggest that there existed, other potential functional uses. Discussions like this produce learning resources that are extremely specific to the question askers' use case.

When helping others use variables, kids would often refer to popular or example projects that contained working code. For instance, K15 asked ``how to make a counter for scores using a sprite making clones of itself''
and was directed by others to an established code chunk ``changeScore method'' in an existing project. In some cases, kids with more advanced knowledge would post snippets of working code:

\begin{quote}
K16: So I have a chat game, when ``hello'' is clicked, the robot would say ``hello''. I wonder how to make the robot say like a option of things such as ``hey'' or ``yo'' instead of ``hello'' all the time?

K17: Put all the hello, hey words in a list then use this code: say (item (random v) of [list v])  %
\end{quote} 

\noindent Although these are wonderful examples of kids mentoring each other, the solutions often offered by kids in our data were so specific---and almost always related to game elements---that they would be unlikely to help a novice learner build a conceptual understanding of data structures. It is not hard to imagine that, if a kid with a specific problem solvable with variables were to browse the discussion threads we analyzed, they might not be able to understand how variables could solve their problem unless they were making a game that was similar to one made by a person who had posted a question.

Furthermore, kids offered code might be able to use it without understanding it \citep{salac_if_2020}. We saw many examples of kids requesting working solutions and many others who seemed happy receiving code that could be copy-and-pasted into their programs. For instance, K18 requested help in the form of an insertable code block: ``Does anyone know how to make a smooth jump script? If you can make it into a custom block that would be great.'' In another example, K19 described the specific effect they wanted and expressed hope that someone could write the code for them:  

\begin{quote}
I want to have a list that has these items: ham, cheese, egg, butter... I need it to find egg and read out it's number in the list. Is there a working script for this?
\end{quote}

\noindent As requested, K19's post was followed by a code snippet with variables named as K19 imagined them.

These examples are part of a broader pattern. To support kids like K18 and K19, the Scratch community creates solutions that are directly applicable to particular use cases. While these responses help kids overcome their problems quickly, directly workable solutions mean that kids might not see the broader picture of how variables and lists can be used.
We explain in the next section how, because learning resources in communities like Scratch consist of questions and solutions accumulated over time, this high degree of specificity in knowledge resources can result in difficult learning experiences for some. 

\subsubsection{Specific examples become archetypes and limit the breadth of functional uses in the community}
\label{sec:restricted_understanding}

Because learning resources were framed around specific examples of functional uses, many of the Scratch users in our forum dataset appeared to have a limited understanding of what variables and lists could do.
This restricted understanding meant that even users without an expressed interest in making games, or particular elements in games, were presented with resources based on them.
For example, a user expressing curiosity about data blocks in very open-ended terms received an answer that based on score counters:

\begin{quote}
K20: I think data blocks can be useful in my projects, but I don't know how to use them.

K21: You are saying variables and lists? For variables, they are just ways to name and store things. So if you make a game and want to keep the score then you'd create a variable called score. %
\end{quote}

\noindent  Although K20 did not express any interest in games, K21's response was focused on them.

This reliance on canonical use cases was particularly obvious in discussions about more advanced cloud variables. For example, in the beginning of the following thread, K22 stated that they did not have knowledge about cloud variables. Despite lack of knowledge on the concept, K22 directly pointed to a canonical use case of it that they had heard of, that is, multiplayer games with leaderboards and high score lists. Immediately after this post, two other kids (K23 and K24) started a discussion thread about the particular use case: 

\begin{quote}
K22: I have no idea what cloud variable is but I heard you could make multiplayer games with it.  

K23: They are shared by 2+ instances of your game. If you make a very simple game in which you add 1 to a cloud variable when a sprite is clicked. Save your game. Then open the game in two new windows in your browser. 

K24: They are usually for High Scores and Multiplayer Games.  
\end{quote}

\noindent Although K22 signalled that they were open to exploration, the answers they received from K23 and K24 indicated the most canonical goals and interests around cloud variables.

We found that kids with interests that deviated from canonical use cases had difficulties finding learning resources that fit their interests. For instance, when K25 posed a very general question about ``how to use cloud variables to save data from users'' in a more general thread about cloud data,
other kids tried to help by posting examples of a high score system that involves cloud variables. K25 expressed confusion because the solution for a high score system did not fit their own goals and said, ``but my game isn't one of those scoring games. I want to make a storyline game.'' 
K25's comment revealed what much of the Scratch forums community takes for granted. K25 ended up not receiving help in the thread. Over time, their post was lost and ignored in a stream of more typical messages about leaderboards.

In summary, we found that certain functional uses became the Scratch forum's go-to examples for explaining variables and lists. Because learning resources were built cumulatively, it is not hard to imagine that more projects with these functional uses of variables and lists would be created over time. These new projects, in turn, became new learning resources as well. Learners who wanted to explore different use cases had fewer relevant resources to guide them.

\subsection{Synthesis: A theory of social feedback loops in interest-driven online learning communities}
\label{sec:theory}

Our findings echo \citepos{papert_mindstorms_1993} emphasis on ``personally powerful ideas'' and \citepos{ito_connected_2013-1} description of interest-driven, community-based learning. We found that kids in Scratch are motivated to use variables and lists to explore their passions and that they leverage content shared by others in the community to do so.  
Because learners are working with specific goals in mind, they run into the need for variables and lists while trying to implement specific elements. They are tutored by peer-produced learning resources framed in terms of those specific functionalities. In this sense, our finding contributes to both the literature of computational participation and computing education by describing that interest-driven content creation can be a potential pathway to support learning of functional uses of complex computing concepts.

At the same time, we also discovered a unintentional side effect of this type of learning. We identified that because community-generated learning resources in the form of Q\&A, tutorials, and project examples tend to be directed toward specific functional uses, those that represent common interests can become archetypes in ways that leave less room for unconventional interests. In some cases, it can also lead to a shallow understanding of underlying concepts \cite{salac_if_2020}. In our sample, the almost exclusive focus on certain game elements raises concerns about whether learners who are not interested in making these elements will be well served by community-generated resources. Furthermore, we observed that kids who were open to exploring different functional uses are pointed to the archetypal use cases. Overtime, this practice can make the most common use cases even more archetypal.

Inspired by the existing body of literature on network-based and social processes that lead to increased concentration of resources over time, such as the Pareto principle~\cite{juran_universals_1954}, preferential attachment~\cite{barabasi_emergence_1999-1}, and the Matthew effect~\cite{merton_matthew_1968}, we theorize that this process plays out as a \textit{social feedback loop}. Our social feedback loop theory is situated in the specific context of interest-driven computational learning in Scratch and can be summarized as follows: \textit{%
interest-driven content creation can result in certain types of creation becoming archetypes that make community-generated learning resources more homogeneous and support an increasingly limited set of learner interests over time.}

\begin{figure*}[t]
  \centering
  \includegraphics[width=0.8\linewidth]{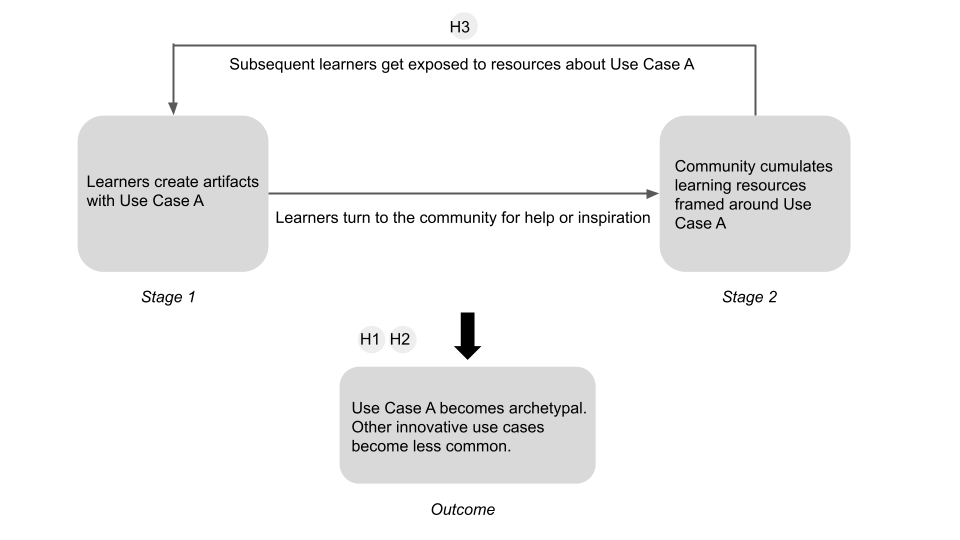}
  \caption{Hypothesized social feedback loop in interest-driven online learning communities}
  \Description{This diagram illustrates the hypothetical social feedback loop that we constructed based on our findings in Study 1. The diagram starts with the box of ``Stage 1'' on the left, and the text that explains Stage 1 says: ``learners create artifacts with Use Case A.'' There is a right-going arrow pointing from Stage 1 to the box of ``Stage 2'' on the right and the text on the arrow says: ``learners turn to the community for help or inspiration.'' The text that explains Study 2 says: ``Community cumulates learning resources framed around Use Case A.'' Above these is a left-going arrow that points from Stage 2 back to Stage 1, forming the loop. The text on the arrow says: ``Subsequent learners get exposed to resources about Use Case A.'' The arrow is also marked with a circle icon with the text ``H3'' in it. Underneath the entire loop there is an down-going arrow pointing to a box of ``Outcome.'' This arrow is marked with two circle icons, same to the one mentioned before, but with ``H1'' and ``H2'' in it. The text that explains Outcome says: ``Use Case A becomes archetypal. Other innovative use cases become less common.''}
  \label{fig:loop}
\end{figure*}

This theory is visualized in Figure \ref{fig:loop}. Stage 1 suggests that some types of creation (Use Case A) will be more popular than others and that more users will tend to create artifacts with Use Case A than other use cases. This may be due to initial user base homogeneity, targeted recruiting in the beginning of the community, examples used in documentation, and so on. 
The arrow pointing from Stage 1 to Stage 2 captures the process of learners running into problems and seeking community support. We argue that users will tend to ask questions framed specifically around Use Case A and draw inspiration from others' artifacts with Use Case A. 
Stage 2 shows the results of this process. As learners successfully receive support, they produce new artifacts with Use Case A that can serve as learning resources for others. %
The arrow on the top of Figure \ref{fig:loop} pointing from Stage 2 back to Stage 1 shows how subsequent learners draw inspiration and support from these accumulated learning resources and, as a result, become even more likely to create artifacts with Use Case A in the future.  
The box on the bottom of Figure \ref{fig:loop} captures the outcome of the feedback loop based on our findings from §\ref{sec:restricted_understanding}. As Use Case A becomes a archetype, the community's collective learning resources are increasingly focused on Use Case A as well. Learners like K25 in §\ref{sec:restricted_understanding} who have different interests receive less support.
Over time, innovative use cases will become less prevalent.

\section{Study 2: Theory testing}
\label{sec:study2}

Study 1's ultimate findings are a set of untested propositions. In a quantitative follow-up study, we conduct tests of three hypotheses that we derived from the theoretical model presented in §\ref{sec:theory} to begin the process of validating the theory. 
Our first two hypotheses (marked as ``H1'' and ``H2'' on the bottom of Figure \ref{fig:loop}) focus on the outcome of the feedback loop, that is, what we will observe if we assume the social feedback loop is occurring. In both cases, the hypotheses are that such a feedback loop will make use cases increasingly homogeneous over time.
First, we hypothesize that certain genres of projects involving simple data structures will become more popular over time relative to other genres. Based on our findings in Study 1, we hypothesize that \textbf{(H1)} \textit{over time, more projects involving variables and lists will be games}. 

Second, we hypothesize that popular functional uses of variables and lists will be even more common relative to others over time. In Scratch, users are able to enter a free form string as the name of their variables and lists. Based on our observations in Study 1, users tend to name variables and lists as the specific things they are trying to make. Therefore, we treated the names that users assign to the variables and lists as a proxy to the functional uses. We thus hypothesize that \textbf{(H2)} \textit{the names that users give to variables and lists will become more concentrated over time}. 

While these hypotheses reflect what we would expect to see in aggregate if the hypothesized social feedback loop were occurring, our third hypothesis (marked as ``H3'' on Figure \ref{fig:loop}) attempts to capture part of the theorized mechanism, that is, users who get exposed to archetype use cases will create similar artifacts.
Therefore, we hypothesize that \textbf{(H3)} \textit{users who have been exposed to projects involving popular variable and list names will be more likely to use those names in their own projects compared to users who have never been exposed to such projects}. 

\subsection{Data}

To conduct our quantitative analyses, we used the \textit{projects}, \textit{project\_s\-trings}, \textit{project\_text} tables from the publicly available Scratch Research Dataset \citep{hill_longitudinal_2017}. For testing H3, we utilized one non-public
column that records which users had downloaded others’ projects. 
We restrict our analysis to projects created between September 2, 2008 and April 10, 2012 because affordances around data blocks were consistent during this period.\footnote{\url{https://en.scratch-wiki.info/wiki/Scratch_1.3}} The period is earlier than the time window used in Study 1 based on differences in the datasets we had access to.
For analytical simplicity, we decided to only include projects with variables and lists written in English.
Finally, we restricted our analysis to \textit{de novo} (i.e., non-remix) projects. 
This resulted in 241,634 projects that contained one or more variables authored by 75,911 Scratch users, and 26,440 projects that contained one or more lists authored by 12,597 users. We created both project-level and user-level datasets with a range of metadata available in the Scratch Research Dataset \cite{hill_longitudinal_2017}. In the spirit of open science, we have placed our full source code for dataset creation and analysis into a public archival repository.\footnote{\url{https://dataverse.harvard.edu/dataset.xhtml?persistentId=doi:10.7910/DVN/2TRZ9N}}

\subsection{Analysis and Measures}

To test \textbf{H1}, we used our project-level dataset to assess whether there is an increase over time in the proportion of games with at least one variable/list.
To ensure that our assumption of games being a predominant genre of project was correct, we randomly subsampled 100 projects with variables and 100 projects with lists. Two coders classified these projects as ``game'' or ``non-game'' and reached high inter-rater reliability (Cohen's $\kappa = 0.88$). We found that 65\% (CI = [54\%, 74\%]) of projects with variables and 52\% (CI = [41\%, 62\%]) with lists were games.\footnote{\label{footnote:ci} Numbers within brackets are 95\% confidence intervals computed using Yates' continuity correction.} This reinforces our sense, developed in Study 1, that games are the dominant genre of Scratch projects containing variables and lists. It also gives us confidence in our decision to use a measure of the prevalence of games over time to test our theory related to popular genres of projects.

Because it is difficult to manually identify games in our large dataset of projects, we define projects as games if they contain the string ``game'' or ``gaming'' in their titles or descriptions. To validate this measure, we again hand-coded samples of projects
from four random samples: 100 projects with variables and at least one of the strings, and 100 similar projects without the strings; two similar samples of 100 projects with lists. The same two coders coded all 400 projects as game or non-games. Among the projects that contain variables, we found that 88\% (CI = [80\%, 93\%]) projects with strings “game” or “gaming” were games, while only 48\% (CI = [38\%, 58\%]) projects without those strings were. We found a similar pattern among projects with lists, where and 85\% (CI = [76\%, 91\%]) and only 31\% (CI = [22\%, 41\%]) projects were games, respectively. In other words, our method of identifying games using the strings ``game'' and ``gaming'' is high precision and somewhat low recall. Because our goal with H1 is to study change over time rather than baseline prevalence, low recall is not problematic as long as it is consistent over time. 
Analysis in our Appendix (§\ref{appendix}) suggests that these proportions were consistent over the period of our study.

To test \textbf{H1}, we perform a logistic regression on the odds of a project with variables/lists being described as a game where the date in years in which the project was created is our independent variable. We include month-of-year fixed effects to control for seasonality. We used a linear specification of time because exploratory data analysis indicated that curvilinearity was unlikely a major concern.

To test \textbf{H2} that there will be increasing concentration in variable/list names over time, we operationalize ``concentration'' as the Gini coefficient of the distributions of variables across names for each week of data we collected  \cite{ceriani_origins_2012}. Originally invented to measure wealth inequality in a nation, Gini coefficients range from 0 representing perfect equality (if every variable name is used in an equal number of projects) to 1 reflecting perfect inequality (if only one variable name were used).
We perform a linear regression using Gini Coefficient as our dependent variable and the same month-of-year fixed effects we use in H1 to control for seasonality. We use a linear specification of time for the same reason we do in H1. 

\textbf{H3} seeks to test the effect of exposure to popular variable/list names on subsequent behavior. We treat popular names as the 20 most frequently used names for variables or lists.
Because it is not possible to measure exposure directly, we use a measure of whether a user has downloaded a project with popular variable/list names as a proxy. We feel this is justified because the only way to access the source code of a Scratch project during our data collection period was to download it.
We use these measures in Cox proportional hazard models \cite{singer_applied_2003}. Originally developed in epidemiology, we follow the framework used by  \citet{dasgupta_remixing_2016} who used Cox models to measure online informal learning of computational thinking concepts in Scratch. Our models  estimate the chance that a user in our dataset will share a \textit{de novo} project with a popular variable name for the first time as a function of the number of \textit{de novo} projects they have previously shared. 

Our question predictor is a time-varying dichotomous measure of whether the user has downloaded a project with a popular variable name during our period of data collection. We conducted the same analysis for lists. Finally, we include a control variable for the total number of downloaded projects to capture overall
exposure to other projects in Scratch---a potential confounder.

\subsection{Results}
\label{sec:study2_results}

\begin{figure*}[t]
     \centering
     \begin{subfigure}[b]{0.48\textwidth}
         \centering
         \includegraphics[width=\textwidth]{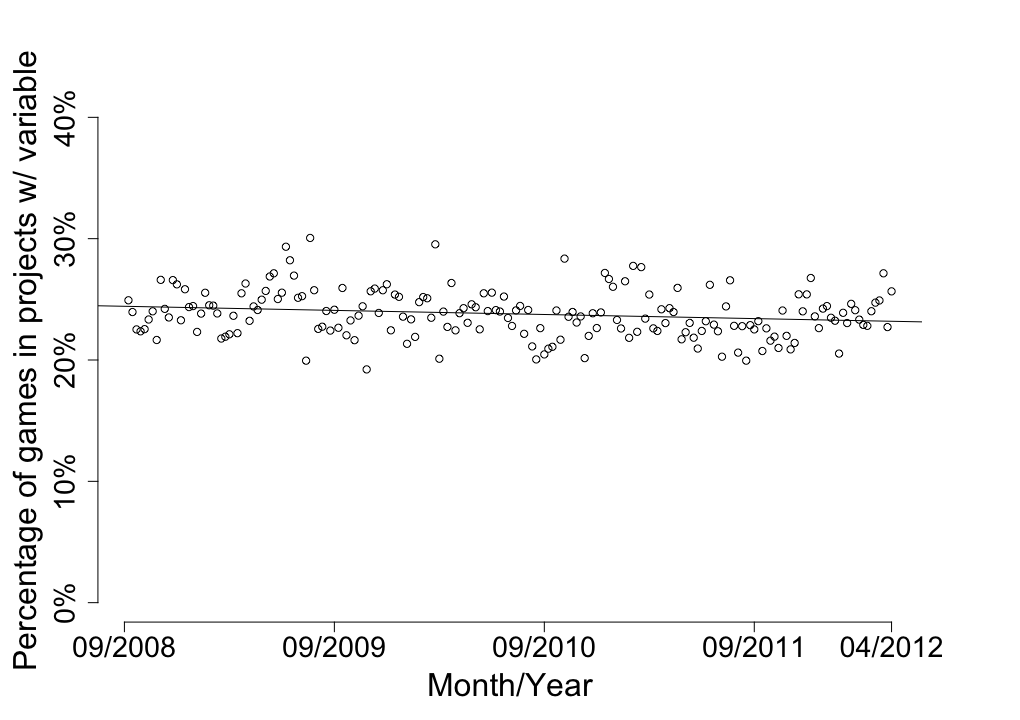}
         \caption{Percentage of games among projects with variables.}
         \Description{This figure is a scatter plot that illustrates the change of percentage of games among projects with variables in the community over time. The x-axis is ``Month/Year.'' The labels are ``09/2008'', ``09/2009'', ``09/2010'', ``09/2011'', and ``04/2012'' from left to right. The y-axis is ``Percentage of games in projects w/variable.'' The labels are ``0\%'', ``10\%'', ``20\%'', ``30\%'', and ``40\%'' from bottom to top. The plotted dots are scattered approximately in the range of 20\% to 30\% on the y-axis across all time period, with a very slight decreasing trend. A bivariate OLS regression line is also plotted, with a intersection with the y-axis at around 25\% and a very slight decreasing trend.}
         \label{fig:game_var}
     \end{subfigure}
     \hfill
     \begin{subfigure}[b]{0.48\textwidth}
         \centering
         \includegraphics[width=\textwidth]{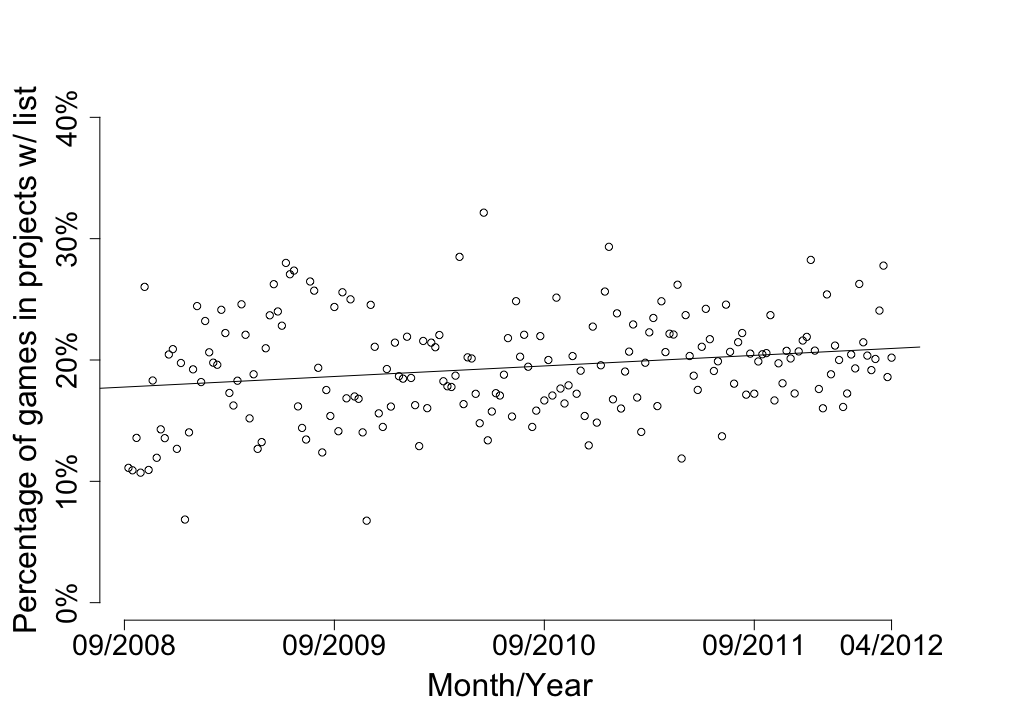}
         \caption{Percentage of games among projects with lists.}
         \Description{This figure is a scatter plot that illustrates the change of percentage of games among projects with lists in the community over time. The x-axis is ``Month/Year.'' The labels are ``09/2008'', ``09/2009'', ``09/2010'', ``09/2011'', and ``04/2012'' from left to right. The y-axis is ``Percentage of games in projects w/variable.'' The labels are ``0\%'', ``10\%'', ``20\%'', ``30\%'', and ``40\%'' from botton to top. The plotted dots are scattered approximately in the range of 10\% to 30\% on the y-axis across all time period, with an increasing trend. A bivariate OLS regression line is also plotted, with a intersection with the y-axis at around 18\% and an increasing trend.}
         \label{fig:game_list}
     \end{subfigure}
     \caption{Percentage of games among projects with variables or lists, per week, from September 2008 to April 2012. Lines reflect bivariate OLS regression lines.}
\end{figure*}

\begin{table}[t]
\begin{center}
\begin{tabular}{l c c}
\hline
 & Variable & List \\
\hline
(Intercept)        & $-1.09^{*}$ & $-1.62^{*}$ \\
                   & $(0.02)$      & $(0.06)$      \\
Year                 & $-0.02^{*}$ & $0.06^{*}$  \\
                   & $(0.00)$      & $(0.02)$      \\
Month fixed effect & yes & yes \\
\hline
Log Likelihood     & $-132558.53$  & $-13058.12$   \\
Deviance           & $265117.06$   & $26116.23$    \\
Num. obs.          & $241634$      & $26440$       \\
\hline
\multicolumn{3}{l}{\scriptsize{$^{*}p<0.001$}}
\end{tabular}
\caption{Logistic regression models for the likelihood of a project including the term ``game'' or ``gaming'' in its title or description. Models are fit on two datasets including all non-remix projects containing variables ($n=241,634$) and all non-remix projects containing lists ($n=26,440$) from September 2008 to April 2012.
}
\label{table:game}
\end{center}
\end{table}

The results from our hypothesis tests provide broad but uneven evidence in support of our theoretical model in §\ref{sec:theory}. 
Figure \ref{fig:game_var} shows that, contrary to H1, the percentage of games in projects with variables decreased slightly over time. The hypothesis test shown in Table \ref{table:game} suggests that this weak relationship is statistically significant ($\beta=-0.02$; $\mathrm{SE}<0.01$; $p<0.01$) and that each year is associated with odds that are 98\% the odds of the year before. On the other hand, our results for lists are in the hypothesized direction. Figure \ref{fig:game_list} shows that the percentage of games among projects with lists has been increasing over time. The results of our logistic regression in Table \ref{table:game} suggest that this relationship is statistically significant ($\beta=0.06$; $\mathrm{SE}=0.02$; $p<0.01$). The model estimates that the odds that a newly created project involving a list is a game are increasing by 106\% year-over-year. For instance, the model-predicted probability of a project with lists created in March 2012 being a game is 22.3\%, while that of a similar project created in March 2009 is 20.8\%.
This estimate translates into 491 more games than we would have expected if there had been no year-over-year increase. We also checked the percentage of games among all projects (not limited to those with variables or lists) over time and found there was no obvious change in the overall game percentage. The details of this analysis is in the Appendix (§\ref{appendix}). In other words, our findings for H1 align with our expectation on the outcome of the hypothetical social feedback loop for lists, but not for variables.

\begin{figure*}[t]
     \centering
     \begin{subfigure}[b]{0.48\textwidth}
         \centering
         \includegraphics[width=\textwidth]{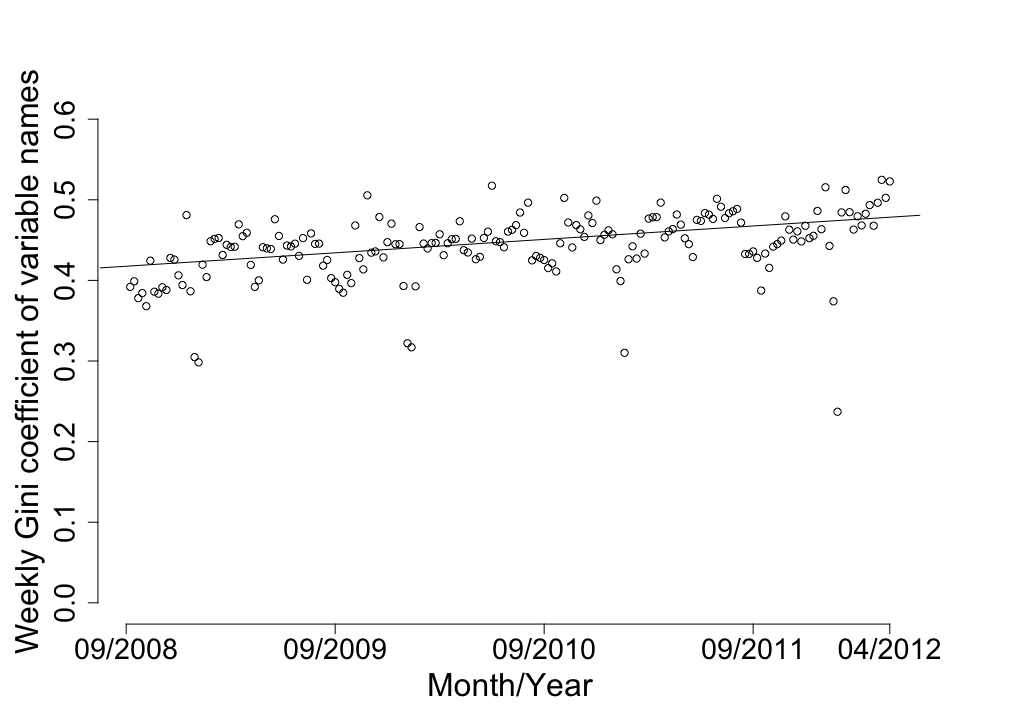}
         \caption{Weekly Gini coefficients of variable names.}
         \Description{This figure is a scatter plot that illustrates the change of weekly Gini coefficient of variable names in the community over time. The x-axis is ``Month/Year.'' The labels are ``09/2008'', ``09/2009'', ``09/2010'', ``09/2011'', and ``04/2012'' from left to right. The y-axis is ``Weekly Gini coefficient of variable names.'' The labels are ``0.0'', ``0.1'', ``0.2'', ``0.3'', ``0.4'', ``0.5'', and ``0.6'' from bottom to top. The plotted dots are scattered approximately in the range of 0.38 to 0.51 on the y-axis across all time period, with an increasing trend. A bivariate OLS regression line is also plotted, with a intersection with the y-axis at around 0.41 and an increasing trend.}
         \label{fig:gini_var}
     \end{subfigure}
     \hfill
     \begin{subfigure}[b]{0.48\textwidth}
         \centering
         \includegraphics[width=\textwidth]{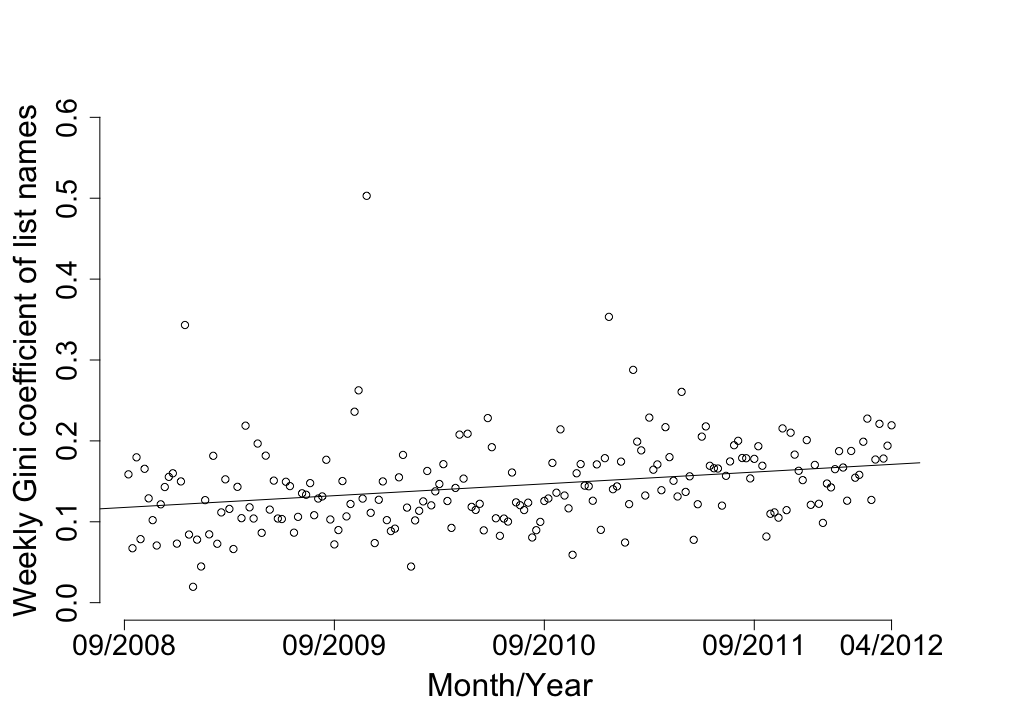}
         \caption{Weekly Gini coefficients of list names.}
         \Description{This figure is a scatter plot that illustrates the change of weekly Gini coefficient of list names in the community over time. The x-axis is ``Month/Year.'' The labels are ``09/2008'', ``09/2009'', ``09/2010'', ``09/2011'', and ``04/2012'' from left to right The y-axis is ``Weekly Gini coefficient of list names.'' The labels are ``0.0'', ``0.1'', ``0.2'', ``0.3'', ``0.4'', ``0.5'', and ``0.6'' from bottom to top. The plotted dots are scattered approximately in the range of 0.05 to 0.22 on the y-axis across all time period, with an increasing trend. A bivariate OLS regression line is also plotted, with a intersection with the y-axis at around 0.11 and an increasing trend.}
         \label{fig:gini_list}
     \end{subfigure}
     \caption{Weekly Gini coefficients of variable and list names over time. Lines reflect bivariate OLS regression lines.}
\end{figure*}

\begin{table}
\begin{center}
\begin{tabular}{l c c}
\hline
 & Variable & List \\
\hline
(Intercept)        & $0.38^{*}$ & $0.11^{*}$ \\
                   & $(0.01)$     & $(0.01)$     \\
Year               & $0.02^{*}$ & $0.01^{*}$ \\
                   & $(0.00)$     & $(0.00)$     \\
Month fixed effect & yes & yes \\
\hline
R$^2$              & $0.38$       & $0.13$       \\
Adj. R$^2$         & $0.34$       & $0.07$       \\
Num. obs.          & $190$        & $190$        \\
\hline
\multicolumn{3}{l}{\scriptsize{$^{*}p<0.001$}}
\end{tabular}
\caption{OLS time series regression models on the Gini coefficient of variables across variables names for all projects shared in Scratch each week ($n=190$).}
\label{table:gini}
\end{center}
\end{table}

We found strong support for H2 that variable and list names would become more concentrated over time.
Figure \ref{fig:gini_var} shows differences in Gini coefficients over time for variables and Figure \ref{fig:gini_list} shows the same measure for lists. Both figures clearly show increasing concentration. Hypothesis tests from OLS time series regression models are reported in Table \ref{table:gini} and reveal that these relationships are statistically significant for both variables ($\beta=0.02$; $\mathrm{SE}<0.01$; $p<0.01$) and lists ($\beta=0.01$; $\mathrm{SE}<0.01$; $p<0.01$). We estimate that the concentration across variables has increased from a Gini coefficient of about 0.41 in 2008 to 0.50 in 2012. 
For reference, this difference is similar to the difference in concentration of wealth between the United States (Gini coefficient = 0.41), which is more concentrated than 68\% of countries globally, and Zimbabwe (Gini coefficient = 0.50) which is more concentrated than 90\%.\footnote{\url{https://data.worldbank.org/indicator/SI.POV.GINI}} 
In other words, the distribution of variables names is both quite concentrated and is increasing in concentration over time. Although list names are much less concentrated in general, they are increasing in concentration at a similar rate. Our findings for H2 provide additional support for what we expect to see in the community if the hypothetical social feedback loop is occurring.

\begin{table*}[!h]
\begin{center}
\begin{tabular}{l c c}
\hline
User & Risk of Using Top & Risk of Using Top \\
& Variable Names & List Names\\
\hline
Downloaded projects w/ & $-0.07^{*}$ & $0.68^{*}$  \\
top variable/list names & $(0.01)$ & $(0.04)$      \\
& & \\
log(\# of 100 downloads)        & $-0.12^{*}$ & $-0.07^{*}$ \\
                        & $(0.02)$ & $(0.02)$      \\
\hline
R$^2$                   & $0.00$ & $0.02$        \\
Max. R$^2$              & $1.00$ & $0.97$        \\
Num. events             & $52967$ & $3790$        \\
Num. obs.               & $88327$ & $17869$       \\
\hline
\multicolumn{2}{l}{\scriptsize{$^{*}p<0.001$; $^{**}p<0.01$; $^{*}p<0.05$}}
\end{tabular}
\caption{Fitted Cox proportional hazard models that estimate the ``risk'' that a Scratch user will share a de novo project that uses a popular variable or list name for the first time, based on whether the user has downloaded a project with top variable or list names. Number of downloads is a control to capture general exposure to other projects in Scratch. A positive coefficient means increased ``risk'', while a negative coefficient means decreased ``risk''.}
\label{table:suv}
\end{center}
\end{table*}

\begin{figure*}[!h]
     \centering
     \begin{subfigure}[b]{0.48\textwidth}
         \centering
         \includegraphics[width=\textwidth]{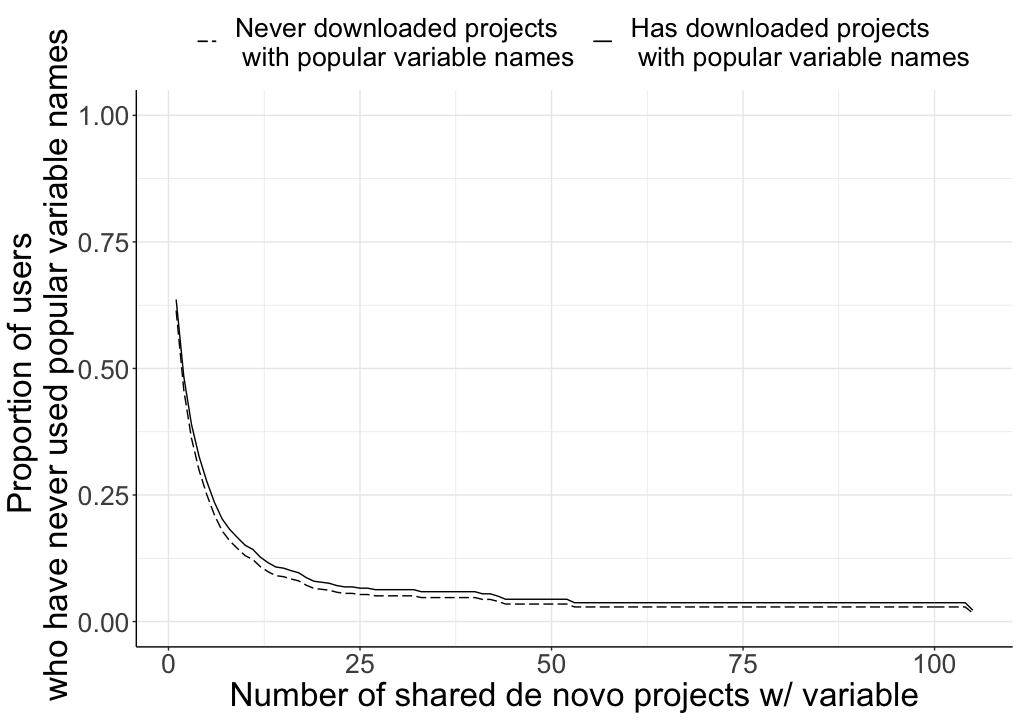}
         \caption{Model-predicted probability of having shared a project with a popular variable name for two prototypical users who have/have not downloaded such a project.
         }
         \Description{This figure is a line plot that illustrates the curves from the survival analysis for variables. The x-axis is ``Number of shared de novo projects w/ variable.'' The labels are ``0'', ``25'', ``50'', ``75'', and ``100'' from left to right. The y-axis is ``Proportion of users who have never used popular variable names.'' The labels are ``0.00'', ``0.25'', ``0.50'', ``0.75'', and ``1.00'' from bottom to top. There are two lines. The dashed line represents users who never downloaded projects with popular variable names. The solid line represents users who has downloaded projects with popular variable names. The solid line starts at 1 on x-axis and approximately 0.63 on the y-axis. The solid line descends steeply in a convex shape and when it reaches 12 on the x-axis, it is at around 0.12 on the y-axis. The line keeps descending, reaches 0.05 on the y-axis when it is at around 40 on the x-axis, and stays at 0.05 for the rest of the x-axis. The dashed line is very closed to the solid line, and it is only slightly more steep than the solid line at the beginning and also stays at 0.05 on the y-axis after reaching 40 at the x-axis.}
         \label{fig:suv_var}
     \end{subfigure}
     \hfill
     \begin{subfigure}[b]{0.48\textwidth}
         \centering
         \includegraphics[width=\textwidth]{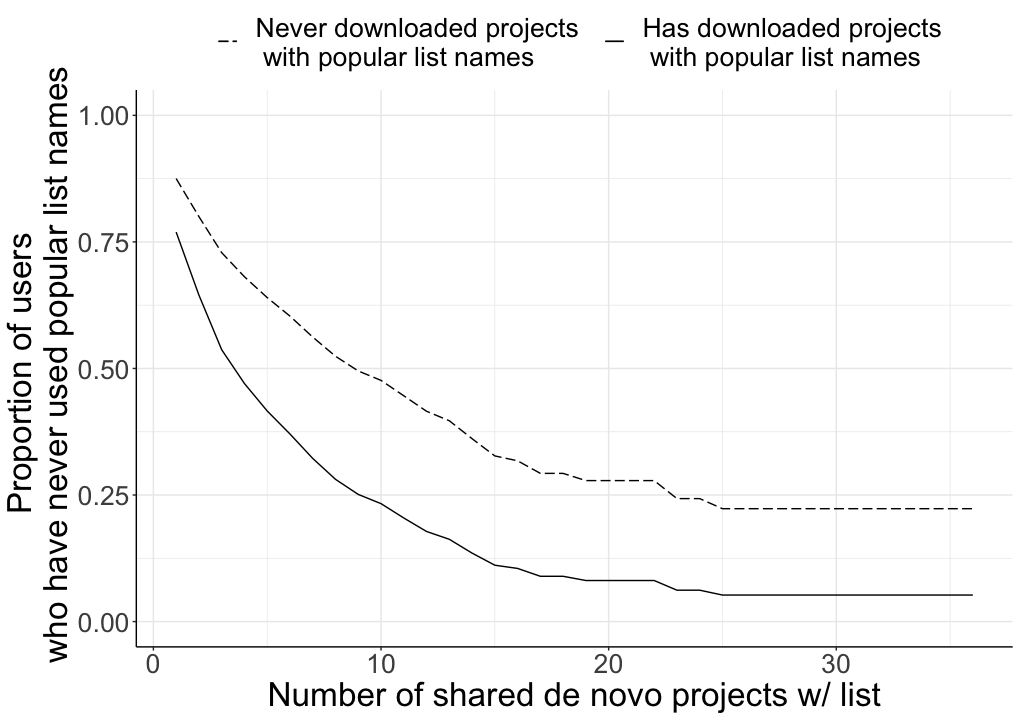}
         \caption{Model-predicted probability of having shared a project with a popular list name for two prototypical users who have/have not downloaded such a project.
         }
         \Description{This figure is a line plot that illustrates the curves from the survival analysis for lists. The x-axis is ``Number of shared de novo projects w/ list.'' The labels are ``0'', ``10'', ``20'', and ``30'' from left to right. The y-axis is ``Proportion of users who have never used popular variable names.'' The labels are ``0.00'', ``0.25'', ``0.50'', ``0.75'', and ``1.00'' from bottom to top. There are two lines. The dashed line represents users who never downloaded projects with popular variable names. The solid line represents users who has downloaded projects with popular variable names.The solid line starts at 1 on x-axis and approximately 0.75 on the y-axis. The solid line descends in a convex shape and when it reaches 10 on the x-axis, it is at around 0.25 on the y-axis. The line keeps descending, reaches around 0.05 on the y-axis when it is at 25 on the x-axis, and stays at 0.05 for the rest of the x-axis. The dashed line is significantly higher than the solid line and stays above it the entire graph. The dashed line starts at 1 on x-axis and approximately 0.88 on the y-axis. The dashed line descends in a convex shape that is less steep than the solid line, and when it reaches 10 on the x-axis, it is at around 0.50 on the y-axis. The line keeps descending, reaches around 0.24 on the y-axis when it is at 25 on the x-axis, and stays at 0.24 for the rest of the x-axis.}
         \label{fig:suv_list}
     \end{subfigure}
     \caption{Plots of model predicted estimates of the proportion for several prototypical users. In Figure (a), estimates are shown for two prototypical users: (dashed) a user who has never downloaded projects with popular variable names, and (solid) a user who has downloaded projects with popular variable names. Figure (b) is the same plot but for lists instead of variables.}
     \label{fig:suv}
\end{figure*}

Table \ref{table:suv} shows parameter estimates from our Cox models and shows mixed support for H3. Although we hypothesized a positive relationship between exposure to and subsequent use of popular variable names, we find that our measure of exposure to popular variables is associated with a risk of using them that is only 93\% as high ($\beta=-0.07$; $\mathrm{SE}=0.01$; $p<=0.001$).
On the other hand, users who downloaded projects with popular list names are more likely to use those names in their own projects than those who did not. 
The instantaneous risk of sharing a project with a popular list name for a user who downloaded at least one project with a popular list name is 1.97 times higher than another similar user who has never downloaded such a project ($\beta=0.68$; $\mathrm{SE}=0.04$; $p<0.001$). The small negative effect associated with variables may be due to the fact that variables are used much more often than lists in Scratch so that kids may simply have more opportunities to be exposed to popular variable names.

Because Cox models are difficult to interpret, we present visualizations of model-predicted estimates in Figure \ref{fig:suv}. Each panel includes two lines reflecting prototypical community members who downloaded one project before sharing projects with others: one prototypical user who had downloaded a project with a popular list or variable name; the other who did not. 
Figure \ref{fig:suv_list} shows that members who have previously downloaded a project with at least one popular list name are more likely to use a popular list name in their own subsequent projects.
Figure \ref{fig:suv_list} shows that our model predicts that \textasciitilde50\% of users who have shared 10 \textit{de novo} projects and who have never downloaded projects with popular list names will have never used a popular name in their own projects, while only \textasciitilde25\% of similar users who have downloaded such projects will not have. Although the negative effect is statistically significant, we do not see a similar effect with variables in Figure \ref{fig:suv_var}. In summary, our findings for H3 offer some evidence for the mechanism of the social feedback loop---one again for lists but not for variables.

\section{Discussion: Challenges \& Opportunities of supporting learning through interest-driven content creation}
\label{sec:discussion}

While researchers have long argued that interest-driven participation can allow learners to explore and be creative \cite{kim_mosaic:_2017, chan_comparing_2016, marlow_rookie_2014, monroy-hernandez_scratchr:_2007}, our case study on computational learning in Scratch indicates that this type of participation might also create self-reinforcing social processes that constrain community imagination around advanced subjects. 
In our first study, we use data from the Scratch forums to build a grounded theory describing a feedback loop that exists between learners' interests and the resources they create. This loop makes it easier for some users to learn about variables and lists---but in ways that are increasingly focused on a set of specific functional uses that have been used by others extensively in the past.
We test several hypotheses derived from this theory in a series of quantitative analyses of the Scratch code corpus and find broad, if uneven, support for the theory.

Our study contributes to the literature on computational participation by highlighting a trade-off between interest-driven participation and learning about computational concepts. On the one hand, our study shows that novice learners can learn functional uses \cite{disessa_models_1986} of advanced computational concepts by engaging in discussion and social support. 
On the other hand, we found that such learning can be superficial and not conceptual or generalizable. Echoing prior studies that raise concern about the lack of depth in the most common forms of computational participation \cite{shorey_hanging_2020, gan_gender_2018}, our study argued that learners' preference for peer-generated learning resources around specific interests can restrict the exploration of broader and more innovative functional uses.   
Although it is conceivable that a narrow set of archetypal use cases could be beneficial for learning for some, increasingly homogeneous use cases stand in clear opposition to the common design goal of broadening participation. 

While our empirical evidence is limited to Scratch, we speculate that our theory describes a common dynamic in informal learning environments. 
In the rest of this section, we discuss three challenges for Scratch and broader online learning communities implied by our theory: (1) a decrease in the diversity of resources that novice learners might draw inspiration from; (2) privileging participation by learners with the most common interests; and (3) a lack of understanding of concepts that cover broad functional use.
We argue that each challenge represents promising opportunities for design.

\subsection{Limited sources of inspiration}

It has been argued that informal learning systems should offer ``wide walls''---affordances that support a range of possibilities and pathways with which learners can creatively construct their own meanings \citep{resnick_reflections_2005, papert_mindstorms_1993}. In the context of Scratch, previous research suggests that novice learners show increased engagement when the walls are ``widened'' through new design features \cite{dasgupta_how_2018}. 
Our findings describe how the unstructured nature of the Scratch online community can lead to overrepresentation of certain ways of applying knowledge---effectively ``narrowing'' the walls.

A range of common social features presented in Scratch and similar online artifact curation and Q\&A communities are likely to reinforce this dynamic. For example, up-votes and likes may externalize the popularity of certain posts \cite{agichtein_finding_2008}, artifact sharing can draw attention to already-visible topics \cite{cheng_building_2020}, and gamified rewards can incentivize already-popular styles \cite{cavusoglu_can_2015}. In each case, these features may make it difficult for learners to see beyond the limited set of popular use cases that the rest of the community is presenting. This narrowing is clearly unintended.
Learners interested in a common application of a concept produce long discussion threads and an abundance of examples and tutorials out of a real desire to help. And indeed, these examples frequently \textit{will} help others. 

Inspired by the call made by \citet{buechley_lilypad_2010}, we suggest that future online informal learning communities should offer affordances that empower learners to leverage the ``long tail'' of novel use cases. Designers of the platforms should seek to help learners recognize new use cases and examples. For instance, designers might highlight novel or unusual projects and provide recognition and status to community members engaged in unconventional creation. For example, the Scratch front page has a curated section designed to showcase projects which could serve this purpose.
Adaptive recommendation systems might help learners broaden their sources of inspiration by directing them to topics and genres that are different from what they are familiar with. Community moderators might guide conversations toward novel perspectives when there has been enough discussion of similar ideas.

\subsection{Narrowed opportunities for participation}

A related challenge is that increasingly homogeneous use cases might marginalize learners not interested in popular topics in ways that lead to demographic inequality.
A number of studies in computational learning communities have shown that the underrepresentation of learners' interests and identity may give rise to a sense of being excluded or marginalized \cite{buechley_lilypad_2010, ford2016paradise, richard2016blind}.
For instance, although many girls use Scratch, there is evidence that girls are generally less interested in making games than boys using the platform \cite{funke_gender_2017, hsu_gender_2014}. As a result, game-specific learning resources related to data structures may make it easier for boys to learn about data.
In that HCI researchers have built curriculum around game-making in Scratch as a way of building up computational thinking \citep{troiano_exploring_2020}, we are concerned about the implications of this approach for users---disproportionately girls---who are not interested in making games.

As a possible step to address this challenge, community designers might elicit users' interests and connect users with similar interests to each other \cite{kraut_building_2011}. The community might also match learners with different backgrounds and interests and motivate them to exchange examples and feedback. Moderators could also offer more support and resources to users who want to explore less popular genres. For example, they might connect users with unusual interests to experts in the community. In the past, the Scratch online community has had a ``welcoming committee'' designed to help newcomers get started \cite{roque_youth_2013}. Our findings suggest a potential way to target these sorts of efforts.

\subsection{Confined understanding of broader knowledge}

The final challenge involves helping learners acquire an understanding of underlying computational concepts that goes beyond specific use cases.
Our findings are consistent with the broader literature on learning and creativity suggesting that when a group of people engage in creative activities, they will generate less diverse ideas after having been shown popular examples \cite{yu_cooks_2011, kulkarni_early_2014}.
Our findings also echo prior work that suggests although the ability to remix can provide inspiration and scaffolds \citep{dasgupta_remixing_2016}, there may be tradeoffs in terms of originality and whether learners might struggle to acquire transformative programming skills \cite{hill_remixing_2013}.
Our study further suggests that although community-produced examples may grow in volume over time, they may only represent material for an increasingly narrow set of functional uses. Informal scaffolds like discussion messages and unregulated artifact catalogs may not always help learners see the big picture or master a skill.

We believe that this challenge points to a final opportunity for learning resource exploration and search systems that focus less on specific examples. For instance, hierarchical tagging and grouping mechanisms could be designed to help novice learners understand the relationship between specific examples and higher level concepts. In Scratch, a high-level collection could be called ``use cases of data structures,'' and the subcategories could include games, story-driven projects, and artistic media. Additionally, the discussion forum could be seeded with prompts to support the identification of underlining conceptual knowledge and to explicitly connect examples with human mentoring \citep{ford_we_2018}, cognitive apprenticeship \cite{hui_introassist_2018} and automatic annotation \cite{chan_comparing_2016}.

\section{Limitations}
First, the interest-driven content creation investigated in this paper is limited to the setup of Scratch forums and projects. For example, Scratch users are mostly children, use block-based programming interface provided by Scratch, and tend to make programs with two-dimensional media. Other computational learning platforms or informal learning communities targeting a different subject may not include these features, and we cannot know exactly how our theory and findings will translate to other contexts. Similarly, since our theory is specifically built around variables and lists, we cannot know how it can be generalized to other type of computational learning and informal learning in general. Although in the discussion section we proposed design implications for online learning communities in general, these implications are merely speculative. Our main contribution is a case study of the Scratch community and we can know when our theories will generalize to other informal learning contexts. We share our work with the hope that future scholars will build on and critique our work by testing these theories in the communities that they study.

Scratch is used in many languages \cite{dasgupta_learning_2017}. Our work is limited in that it only considers English language content. 
We do not know what impact the multi-lingual nature of Scratch has on our analysis or if the dynamics we observe are also present in other linguistic subcommunities in Scratch. Our strategy to detect project genre in our test of H1 is limited by language in that not all games have the words ``game'' or ``gaming'' in their title or description and some non-game projects do.
Additionally, Scratch users learn from resources including project comments, tutorials, and one-to-one mentorship both within and beyond the Scratch community. In that forums are not the sole (or even primary) way that kids learn in Scratch, Study 1 might be missing important social dynamics in other places. 

As we discussed in §\ref{sec:study1}, our sample in Study 1 is nonrepresentative in ways that may shape our findings. Because a quarter of our sample selects on the 11 most popular variable/list names---and because these names mostly indicate game elements---our qualitative dataset may be skewed toward game-making in ways that shape our findings in Study 1. The random samples and population-level data used on Study 2 is an effort to address this issue.

Another set of limitations stems from our reliance on imperfect proxy measures in Study 2. For example, we use downloads as a measure of exposure to test H3 because downloading was the only way to view the source code of a Scratch project before 2013. That said, users might download projects to deal with a slow internet connection or for a range of other reasons.\footnote{\url{https://en.scratch-wiki.info/wiki/Project_Downloading\#Benefits_of_Downloading_Projects}} 
Although we feel confident that downloads will be correlated with exposure, we have no way of knowing why a user downloaded any given project.
Similarly, we used user-defined names of variables and lists as a proxy of the use cases of Scratch data structures. Although our findings from Study 1 suggest that variable and list names largely represent what users were making with variables and lists, we cannot know for sure if this is accurate in every project and there is chance that some users may use names that are inconsistent with the use case. In addition, like most other studies of informal learning online, we can only observe learning experiences and not outcomes. Measuring learning outcomes in a community like Scratch is difficult because learners arrive with different interests and aspirations and take different paths.
Although we measure the presence or absence of variables and lists in projects, we can not know whether they are being used correctly or whether project creators understand the code they write \cite{salac_if_2020}.

Finally, although we theorize that there is a causal link between our proposed social feedback loop and increased homogenization of community-produced learning resources, we present no causal evidence. For example, our test of H3 provides evidence of a correlation between exposure to popular list names and an increased likelihood of future use of those names. This relationship might also be due to variables that are correlated with, but not caused by, a social feedback loop like the one we describe. Similarly, we try to argue that the narrowing trend in the usage of variable and list names that we discovered is an indicator of the social feedback loop that narrows the creativity in the community. However, there may be other factors outside the community, such as pop-cultural trends and school education, that contribute to this narrowing effect. To summarize, the evidence we present in Study 2 should be interpreted as similar to what we would expect to find if our theory were true---nothing more. 

\section{Conclusion}

Our work contributes to HCI and social computing research by presenting a mixed-method case study about how users learn data structures through interest-driven content creation in the Scratch online community. We found evidence of a problematic, but previously untheorized, feedback loop that can constrain community innovation of functional uses. %
Our work describes how informal learning communities similar to Scratch could find it increasingly difficult to serve users who interests are outside the mainstream of their communities. Most importantly, our work points to several promising paths forward for designers of these systems. The problems we highlight reflect several ways that informal online learning communities could more effectively realize their incredible promise. We hope our work furthers this goal.

\begin{acks}
We would like to acknowledge members of the Community Data Science Collective---especially Emilia Gan and Stefania Druga---for their feedback and support. We also want to thank our anonymous reviewers for their feedback and support.
We want to thank the Scratch team---especially Mitchel Resnick,  Natalie Rusk, and Andrés Monroy-Hernandez---for building the Scratch online communities and for making data about Scratch available that allowed us to conduct this work.
Finally, we owe a huge debt of gratitude to members of the Scratch online community who continue to inspire and impress us and without whom this work would not be possible.
\end{acks}

\bibliographystyle{ACM-Reference-Format}
\bibliography{references}

\newpage
\section{Appendix}
\label{appendix}
\appendix

\begin{figure}[!h]
  \centering
  \includegraphics[width=1\linewidth]{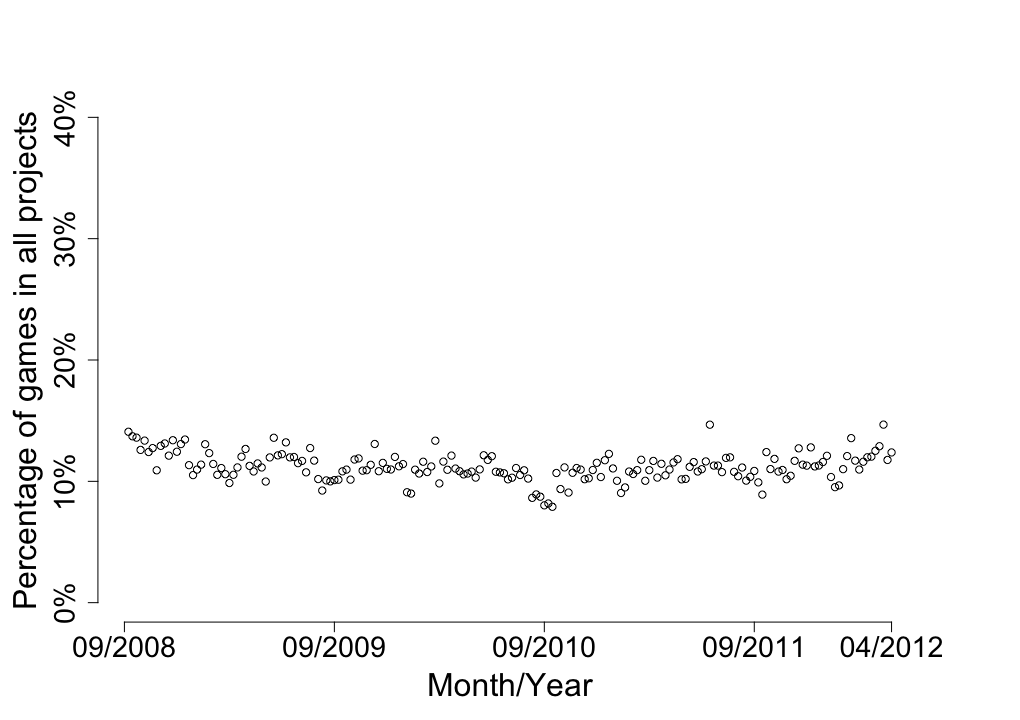}
  \caption{Percentage of games among all projects in the community, per week, from September 2008 to April 2012.}
  \Description{This figure is a scatter plot that illustrates the change of percentage of games among all projects in the community over time. The x-axis is ``Month/Year.'' The labels are ``09/2008'', ``09/2009'', ``09/2010'', ``09/2011'', and ``04/2012'' from left to right. The y-axis is ``Percentage of games in projects w/variable.'' The labels are ``0\%'', ``10\%'', ``20\%'', ``30\%'', and ``40\%'' from bottom to top. The plotted dots are scattered approximately in the range of 10\% to 15\% on the y-axis across all time period, with a flat trend.}
  \label{fig:game_overall}
\end{figure}

Figure \ref{fig:game_overall} shows the trend of percentage of games among all projects (not limited to those with variables or lists) over time as part of the analysis for H1 in §\ref{sec:study2_results}. As shown in the figure, the overall percentage of games stayed consistent in the time period of our study.

To get a sense of whether the precision and recall of our strategy to identify games stayed consistent over time, we calculated the precision and recall in the first (09/02/2008-06/22/2010) and second half (06/22/2010-04/10/2012) of our study time period, for projects with lists and projects with variables respectively. The precisions were 0.83 and 0.88 for projects with lists and 0.95 and 0.83 for projects with variables. The recalls were 0.60 and 0.62 for projects with lists and 0.56 and 0.56 for projects with variables. This means the precision and recall largely stayed consistent.

\end{document}